\documentclass[article]{revtex4}
\usepackage{graphicx}
\usepackage{dcolumn}
\usepackage{amsmath}
\usepackage[latin1]{inputenc}
\usepackage{graphicx, psfrag}
\usepackage{amssymb}
\usepackage[colorlinks=true, citecolor=blue, urlcolor = blue, linkcolor= red, bookmarks=true]{hyperref}
\usepackage{float}
\usepackage{amsmath}
\usepackage{amsfonts}
\usepackage{dcolumn}
\usepackage{hyperref}
\usepackage{subfigure}
\usepackage{pgfplots}
\usepackage{epstopdf}
\usepackage{booktabs}

\makeatletter
\def\btt#1{\texttt{\@backslashchar#1}}
\DeclareRobustCommand\bblash{\btt{\@backslashchar}} \makeatother

\begin{document}
\title{Rotating charged black hole shadow in quintessence }
\author{Balendra Pratap Singh}
\email{balendra29@gmail.com}
\affiliation{Department of Applied Sciences and Engineering,\\ Tula's Institute, Dehradun, Uttarakhand, India.}
\begin{abstract}
 We analytically  study the shadow of the charged rotating black hole in the presence of quintessence. The quintessential energy adequately affects the shape and size of the black hole shadow. The shadow of a rotating black hole spots a distorted dark disk. The shape and size of the black hole shadow depend on its mass $M$, spin parameter $a$, charge $q$,  quintessential field parameter $\omega_q$, and normalization factor $c$.  We derive the entire geodesic structure of photons near black holes using the Hamilton-Jacobi equation and Carter constant separable method. We relate the celestial coordinates to geodesic equations and plot the contour of the black hole shadow for the case $\omega_q=-2/3$. To find the effect of quintessential energy on the black hole shadow, we obtain observable  i.e. shadow radius $R_s$ and distortion parameter $\delta_s$.  We compare all our results with the Kerr black hole in quintessence, Kerr-Newman black hole and Bardeen--Kiselev black hole. Our study shows that for a fixed value of the spin parameter $a$ and normalization factor $c$, the black hole shadow monotonically decreases and gets more distorted with charge $q$.  
\end{abstract}
\maketitle
\section{Introduction}
The event horizon telescope (ETH) association has published the first photograph of a supermassive black hole at the center of the galaxy Messier 87 (M87)\cite{Akiyama:2019cqa}-\cite{Akiyama:2019eap}. The observed image of the black hole is a dark spot surrounded by bright  rings. The dark region is called the black hole shadow and the bright rings are the photon rings. A black hole is in itself invisible whereas it casts a shadow due to the deflection of light in its strong gravitational field.  The image of the black hole is another strong evidence of the confirmation of Einstein's general relativity and opens a new gateway to test alternative theories of gravity.

  The very first study of deflection of light around Schwarzschild black hole was done by Synge \cite{Synge:1966} and after that Luminet constructed a simulated photograph of the shadow of Schwarzschild black hole \cite{Luminet:1979}.
    The shadow cast by rotating black hole was discussed by Bardeen \cite{Bardeen:1973gb}.  The black hole shadow appears distorted disc for a rotating spacetime. The apparent shapes of the other rotating spacetimes studied by several authors, such as Kerr-Newman  \cite{Takahashi:2005hy}, Einstein--Maxwell--Dilation--Axion \cite{Wei:2013kza}, Kerr-Taub-NUT \cite{Abdujabbarov:2012bn}, Braneworld \cite{Amarilla:2011fx}, Kaluza--Klein rotating dilation \cite{Amarilla:2013sj}, Non-Kerr \cite{Atamurotov:2013sca}, Konoplya--Zhidenko rotating non-Kerr \cite{Wang:2017hjl}, Sen \cite{sen:2009}, Kerr--Newman--Nut with cosmological constant \cite{Grenzebach:2014fha}, nonsingular \cite{Amir:2016cen}, regular black holes \cite{Abdujabbarov:2016hnw}, asymptotically safe gravity \cite{Kumar:2019ohr}, rastall theory of gravity \cite{Kumar:2017tdw}  and $4D$ Einstein--Gauss--Bonnet gravity \cite{Kumar:2020owy}. 
The possibility of finding a black hole in front of a bright object is very small, so for the regular observation Ref.~\cite{Atamurotov:2015nra} and \cite{Perlick:2015vta} discussed the optical properties of black holes in the presence of plasma \cite{Goddi:2016jrs}. The shadow of the Kerr black hole with scalar hair and without scalar hair has been investigated by \cite{Cunha:2015yba,Cunha:2016bpi}.
 To estimate the black hole parameters from its shadow images, several researchers proposed different methods to find observables \cite{Abdujabbarov:2015xqa,Hioki:2009na,Kumar:2018ple,Chang:2020miq,Pesce:2021adg} .
This subject extended to higher--dimensional spacetime by several researchers \cite{Papnoi:2014aaa,Abdujabbarov:2015rqa,Amir:2017slq,Singh:2017vfr,Belhaj:2020rdb}. 
The cosmological observations have confirmed the accelerating expansion of the Universe \cite{SupernovaSearchTeam:2004lze,Peebles:2002gy,Copeland:2006wr,WMAP:2010qai}. The expansion of the Universe requires the existence of dark energy. 
Quintessence is a dark energy candidate which sufficiently affects the spacetime structure of the black hole \cite{Kiselev:2002dx}. Recently several authors studied the  shadow images with quintessential energy for different spacetimes \cite{Abdujabbarov:2015pqp,Belhaj2022,Zeng:2020vsj,Sun:2022wya,Pedraza:2020uuy}. Motivated from the study of rotating black hole shadow with the quintessential energy \cite{Abdujabbarov:2015pqp} we study the effect of charge for rotating spacetime in quintessence.

We organized this paper as follows: In Sect.~\ref{sect2} we describe the properties of  non-rotating and rotating  black hole metric. In  Sect.~\ref{sect3} we study the motion of a test particle and derive its complete null geodesic equations by using the Hamilton-Jacobi and Carter separable method. For the visualization of the black hole shadow, we derive celestial coordinates in Sect.~\ref{sect4} and plot the contour of the black hole shadow. We also study the observables: shadow radius and distortion parameter from the contour of black hole shadow in Sect.~\ref{sect4}. We conclude our results in Sect.~\ref{sect5} .
\section{Charged black hole in quintessence}\label{sect2}
In this section, we discuss the black hole metric for non-rotating and rotating cases. 
\subsection{Non-Rotating charged black hole in quintessence}
A black hole with zero angular momentum is considered  a non-rotating black hole.  The static spherically symmetric charged black hole metric in quintessence  can be expressed by the following equation\cite{Kiselev:2002dx}
\begin{equation}
ds^{2}=-f(r)dt^{2}+\frac{1}{f(r)}dr^{2}+r^{2}d\Omega^{2},
\label{q1}
\end{equation}
with 
\begin{equation}\label{f(r)}
f(r)=1-\frac{2M}{r}+\frac{q^{2}}{r^{2}}-\frac{c}{r^{3\omega_q+1}},
\end{equation}
where  $M$ is the black hole mass, $q$ is the black hole charge,   $c$ is the normalization factor and $\omega_q$ is the quintessential field parameter. The normalization factor gives the intensity of the quintessential field around the black hole and quintessential field parameter describes the equation of state. The density of quintessence matter $\rho_{q}$ is related to the normalization factor $c$ via
\begin{equation}
{\rho_q} =-\frac{c}{2}\frac{3\omega_{q}}{r^3(\omega_q+1)}.
\end{equation}   
The quintessence matter density is ${\rho_q}>0$. Hence, the normalization factor has  $+$ve values ($c>0$) for the $-$ve values of state parameter ($\omega_q<0$). The quintessential state parameter determines the properties of the black hole metric.  The parameter $\omega_q$ can take the value between $-1<\omega_q<-1/3$ and related with pressure $p$ and density $\rho_q$ by the equation of state $p=\omega_q \rho_q$. For $-1/3<\omega_q<0$ ,the black hole metric has the asymptotically flat solution. 
The de sitter horizons exist for $-1<\omega_q<-1/3$. In the absence of quintessential field $c=0$, the black hole metric reduces to Reissner-Nordstrom spacetime. 
\subsection{Rotating charged black hole with  quintessence}
The rotating part of the hole metric with  quintessence can be obtained by applying the Newman--Janis algorithm \cite{Newman-Janis:1965} to the non-rotating  black hole with quintessence. In the Boyer-Lindquist co-ordinate the metric takes the form  \cite{Xu:2016jod,Azreg-Ainou:2014pra}, 
\begin{eqnarray}
ds^{2}&=&-\left(1-\frac{2Mr-q^2+cr^{1-3\omega_{q}}}{\Sigma}\right)dt^{2}+\frac{\Sigma}{\Delta}dr^2-\frac{2a\sin^{2}\theta(2Mr-q^2+cr^{1-3\omega_{q}})}{\Sigma}  \nonumber \\ \nonumber
&+&\sin^{2}\theta\left(r^2+a^2+a^2\sin^2{\theta}\frac{2Mr-q^2-cr^{1-3\omega_q}}{\Sigma}\right)d\phi^{2}
\label{16}
\end{eqnarray}
where
\begin{eqnarray}
\Delta &=&r^{2}-2Mr+a^{2}+q^{2}-c r^{1-3\omega_q},\\ \label{17}
{\Sigma}^2&=&r^2+a^2\cos^{2}{\theta}.
\end{eqnarray}
In the absence of charge $q=0$, the  black hole metric  reduces to Kerr black hole in quintessence \cite{Ghosh:2015ovj,Toshmatov:2015npp} also in the limit $c\rightarrow0$ the black hole metric reduces to Kerr black hole.

\section{Null Geodesics and impact parameters} \label{sect3}

In this section, we derive the geodesic equations of motion for photons. We adopt the Hamilton-Jacobi equation and Carter constant separable method \cite{Carter:1968rr} to study the complete geodesic structure of photons near the black hole. In our study, we considered a fixed value, of the quintessential parameter $\omega_q=-2/3$ and  for this value, the black hole metric reduces to the following form
\begin{eqnarray}\label{metric}
ds^2&=&-\left(1-\frac{2\varrho r}{\Sigma}\right)dt^2+\frac{\Sigma}{\Delta}dr^2-\frac{4a\varrho r\sin^2\theta}{\Sigma}d\phi dt\nonumber\\&&+\Sigma d\theta^2 +\sin^2\theta\left(r^2+a^2+a^2\sin^2\theta\frac{2\varrho r}{\Sigma}\right)d\phi^2 ,
\end{eqnarray}
 where
\begin{eqnarray} \label{metricfuncts}
\Delta&=&r^2-2\varrho r+a^2, \nonumber\\
\varrho&=&M+\frac{ c}{2}r^2-\frac{q^2}{2r} , \nonumber\\
\Sigma&=& r^2+a^2 \cos^2\theta. \nonumber
\end{eqnarray}
The most general form of the Hamilton-Jacobi equation  reads
\begin{eqnarray}
\label{HmaJam}
\frac{\partial S}{\partial \sigma} = -\frac{1}{2}g^{\alpha\beta} p_\alpha p_\beta ,
\end{eqnarray}
where $p_\alpha= \partial S/\partial x^\alpha$,  and $S$ is the Jacobean action. The separable solution for the Jacobi action $S$  can be chosen as
\begin{eqnarray}
S=\frac12 {m_0}^2 \sigma -{\cal E} t +{\cal L} \phi +S_r(r)+S_\theta(\theta) ,
\end{eqnarray}
where $m_0$ is the mass of the test particle, which is zero in the case of photons. In the black hole metric (\ref{metric}) the coordinates $t$, $\phi$ are cyclic and corresponding to these cyclic coordinates two conserved quantities energy $\mathcal{E}$ and angular momentum $\mathcal {L}$ exist. Using the variable separable method, we obtain the complete equations of motion for photons around the  black hole,  which takes the following form
\begin{eqnarray}
\Sigma \frac{dt}{d\sigma}&=&\frac{r^2+a^2}{\Delta}\left[{\cal E}(r^2+a^2)-a{\cal L}\right]  -a(a{\cal E}\sin^2\theta-{\mathcal {L}})\ ,\label{t}\\
\Sigma \frac{dr}{d\sigma}&=&\sqrt{\mathcal{R}(r)}\ ,\label{r}\\
\Sigma \frac{d\theta}{d\sigma}&=&\sqrt{\Theta(\theta)}\ ,\label{th}\\
\Sigma \frac{d\phi}{d\sigma}&=&\frac{a}{\Delta}\left[{\cal E}(r^2+a^2)-a{\cal L}\right]-\left(a{\cal E}-\frac{{\cal L}}{\sin^2\theta}\right)\ .\label{phi}
\end{eqnarray}
where the expressions for 
$\mathcal{R} (r)$ and ${\Theta}(\theta)$ in Eq.~(\ref{r}) and (\ref{th}) are
\begin{eqnarray}\label{06}
&&\mathcal{R}(r)=\left[(r^2+a^2){\cal E}-a{\cal L}\right]^2-\Delta\left[{m_0}^2r^2+(a{\cal E}-{\cal L})^2+{\cal K}\right],\\
&&\Theta(\theta)={\cal K}-\left[\frac{{\cal L}^2}{\sin^2\theta}-a^2 {\cal E}^2\right]\cos^2\theta\ ,
\end{eqnarray}
with separability Carter constant  $\mathcal{K}$. The dynamics of the test particle around rotating charge black holes in quintessence are completely determined by the Eqs.~(\ref{t})-(\ref{phi}). In the limit of $q\rightarrow0$, these geodesic equations reduce for Kerr black hole in quintessence \cite{Abdujabbarov:2015pqp} and when $c=0$, these geodesic equations reduce for Kerr black hole \cite{Frolov:1998wf}. 

Next, we are willing to study the geometry of photons near the black hole. We are introducing two impact parameters $\eta$ and $\xi$  in terms of energy $\mathcal{E}$, angular momentum $\mathcal{L}$, and carter constant $\mathcal{K}$
\begin{equation}
\xi=\mathcal{L}/\mathcal{E}, \quad\quad \eta=\mathcal{K}/\mathcal{E}^2.
\end{equation}
One can find the radial equation of motion for photons in terms of dimensional quantities $\eta$ and $\xi$ as
\begin{equation}
\mathcal{R}(r)=\frac{1}{{\cal{E}}^2}\left[(r^2+a^2) -a{\xi}\right]^2-\Delta\left[(a -{\xi})^2+{\eta}\right].
\end{equation}
The radial equation of motion can be rewritten in terms of the effective potential $V_{eff}$ as
\begin{equation}
\left(\frac{d{r}}{d\tau}\right)^2+V_{eff}(r)=0,
\end{equation}
where the effective potential $V_{eff}$ takes the form 
\begin{equation}
V_{eff}=\frac{1}{{{\Sigma}^2}}\left[(r^2+a^2) -a{\xi}\right]^2-\Delta\left[(a -{\xi})^2+{\eta}\right].\label{vef}
\end{equation}
We can find the most critical and unstable circular orbits by maximizing the effective potential, which satisfies the condition
\begin{equation}
V_{eff}=\frac{\partial V_{eff}}{\partial r}=0 \;\; \;\; \mbox{or}\;\;  \; \mathcal{R}=\frac{\partial \mathcal{R}}{\partial r}=0.\label{vr} 
\end{equation}
one can find the equations for  impact parameters $\eta$ and $\xi$ by applying the condition mentioned in Eq.~(\ref{vr}) 
\begin{eqnarray}
\xi &=& \frac{a^2 (r (2-3 c r)+2 M)+r \left(r (r (c r+2)-6 M)+4 q^2\right)}{a (2 M-r (3 c r+2))} \ , \label{xiexp}\\
\eta&=&  -\frac{r^2 \left(8 a^2 \left(-c r^3-2 M r+2 q^2\right)+\left(r (r (c r+2)-6 M)+4 q^2\right)^2\right)}{a^2 (r (3 c r+2)-2 M)^2}. \label{etaexp}
\end{eqnarray}
The expressions  of $\xi$  and $\eta$  in Eqn. (\ref{xiexp}) and (\ref{etaexp})  can define the complete apparent shape of the rotating charged black hole in quintessence which may refer to the  Kerr black hole  in quintessence when $q=0$ \cite{Abdujabbarov:2015pqp} and Kerr black hole in the limit of $c=0$ \cite{Frolov:1998wf}.
\section{Quantitative analysis of the black hole shadow}  \label{sect4}
\begin{figure*}
    \begin{tabular}{c c c c}
	\includegraphics[scale=0.6]{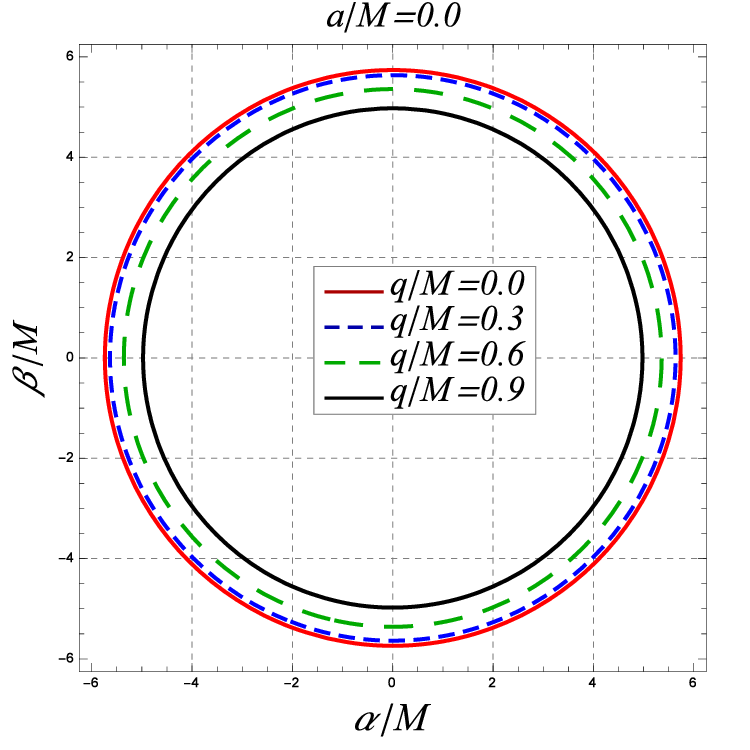} 
	\includegraphics[scale=0.6]{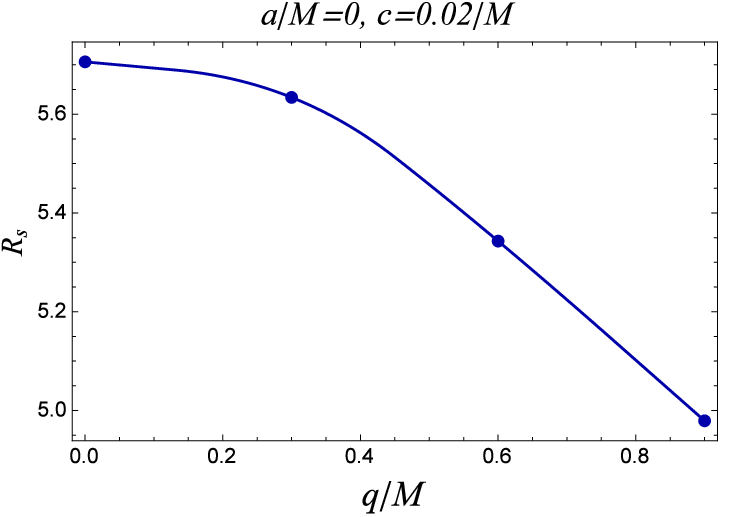} 
	 \end{tabular}
    \caption{\label{Fig1} Plot shows the shadows of non-rotating charged black hole in quintessence for fixed values of $c=0.02$ with charge $q$.}
\end{figure*}
We have calculated the impact parameters which trace the photon orbits around the black hole. For visualization of the black hole's shadow, we require celestial coordinates. The visual image of the black hole's shadow is a dark spot surrounded by bright photon rings. The incoming photons with larger angular momentum fall inside the black hole and spot a dark circle, while photons with smaller angular momentum turn back on a certain turning point or scatter from the black hole are observed by an observer at infinity. Here, we study the in falling null geodesics and for that, we are introducing two celestial coordinates, $\alpha$, and $\beta$
\begin{figure*}
    \begin{tabular}{c c c c}
	\includegraphics[scale=0.5]{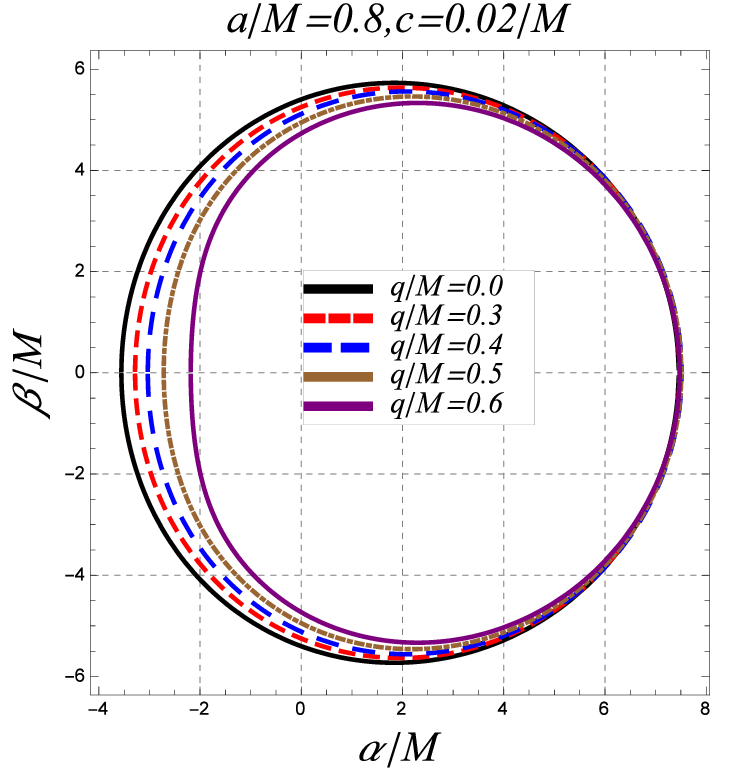} 
	\includegraphics[scale=0.5]{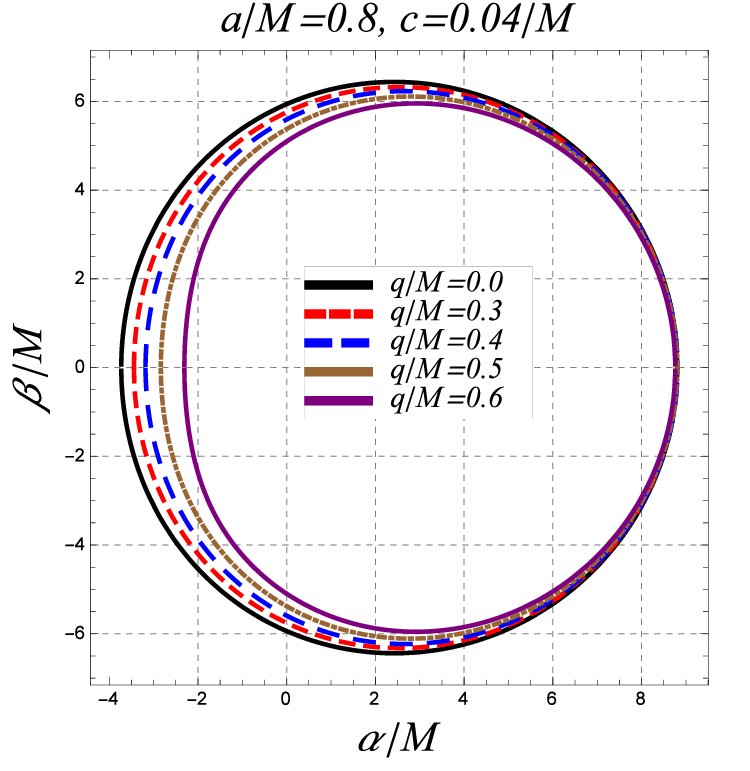} \\
	\includegraphics[scale=0.5]{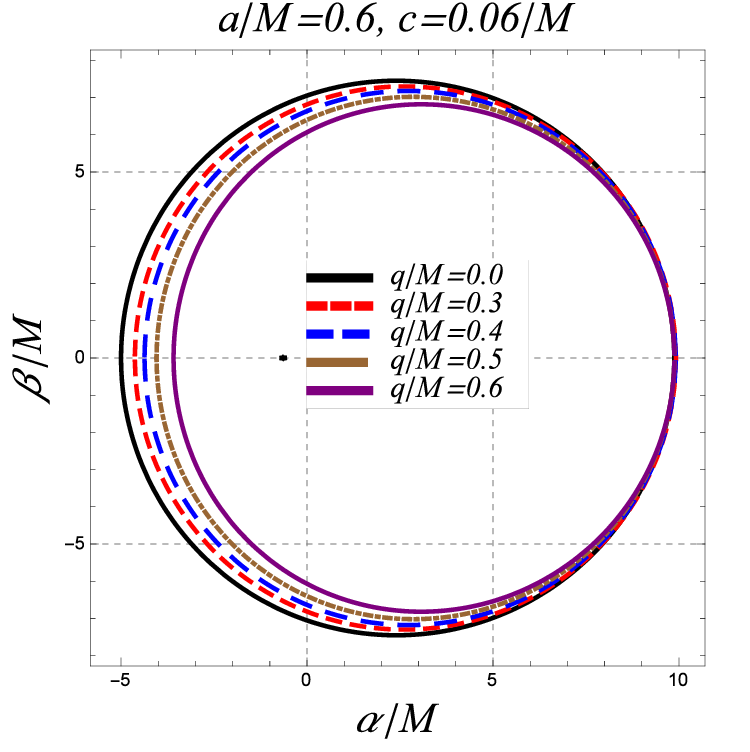} 
	\includegraphics[scale=0.5]{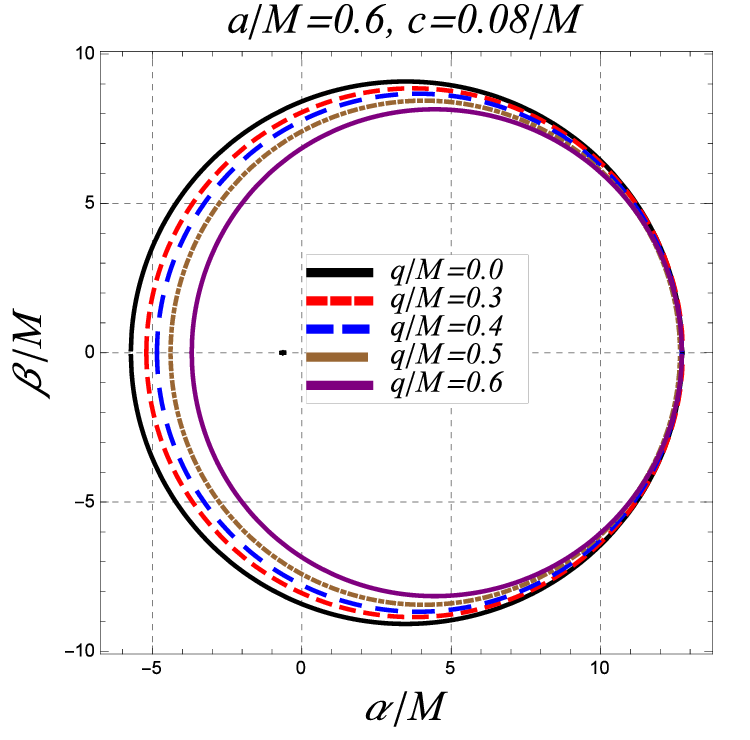} 
	 \end{tabular}
    \caption{\label{Fig2} Shadows of  rotating charged black hole in quintessence for different values of spin parameter $a$, quintessence field parameter $c$ and charge $q$.}
    \end{figure*}
\begin{eqnarray}
&& \alpha=\lim_{r_0\rightarrow\infty}\left(-r_0^2 \sin{\theta_0}\frac{d\phi}{d{t}}\right),\\
&& \beta=\lim_{r_0\rightarrow\infty}r_0^2\frac{d\theta}{dr},
\end{eqnarray}
where $r_0$ is the distance between the black hole and the far distance observer, $\theta_0$ is the inclination angle between the black hole rotation axis and line of sight from source to observer. From null geodesic Eqs.~(\ref{t})-(\ref{phi}), we find the relation between celestial coordinates and constants $\eta$ and $\xi$ as
 \begin{eqnarray}
\alpha&=& -\frac{\xi}{\sin\theta}\, \label{alpha}\ ,\\
\beta&=&\sqrt{\eta+a^2\cos^2\theta-\xi^2\cot^2\theta } \label{beta}\ .
\end{eqnarray}
  Eqs.~(\ref{alpha}) and (\ref{beta}) show the relation between celestial coordinate $\alpha$, $\beta$ to impact parameter  $\eta$ and $\xi$. The equations of celestial coordinates at the equatorial plane ($\theta_0$ =$\pi/2$) reduce to
\begin{eqnarray} 
&&\alpha=-\xi, \label{le} \\ 
&& \beta=\pm\sqrt{\eta}.\label{pt}
\end{eqnarray}
\begin{figure*}
\begin{center}
     \begin{tabular}{c c c c}
	\includegraphics[scale=0.35]{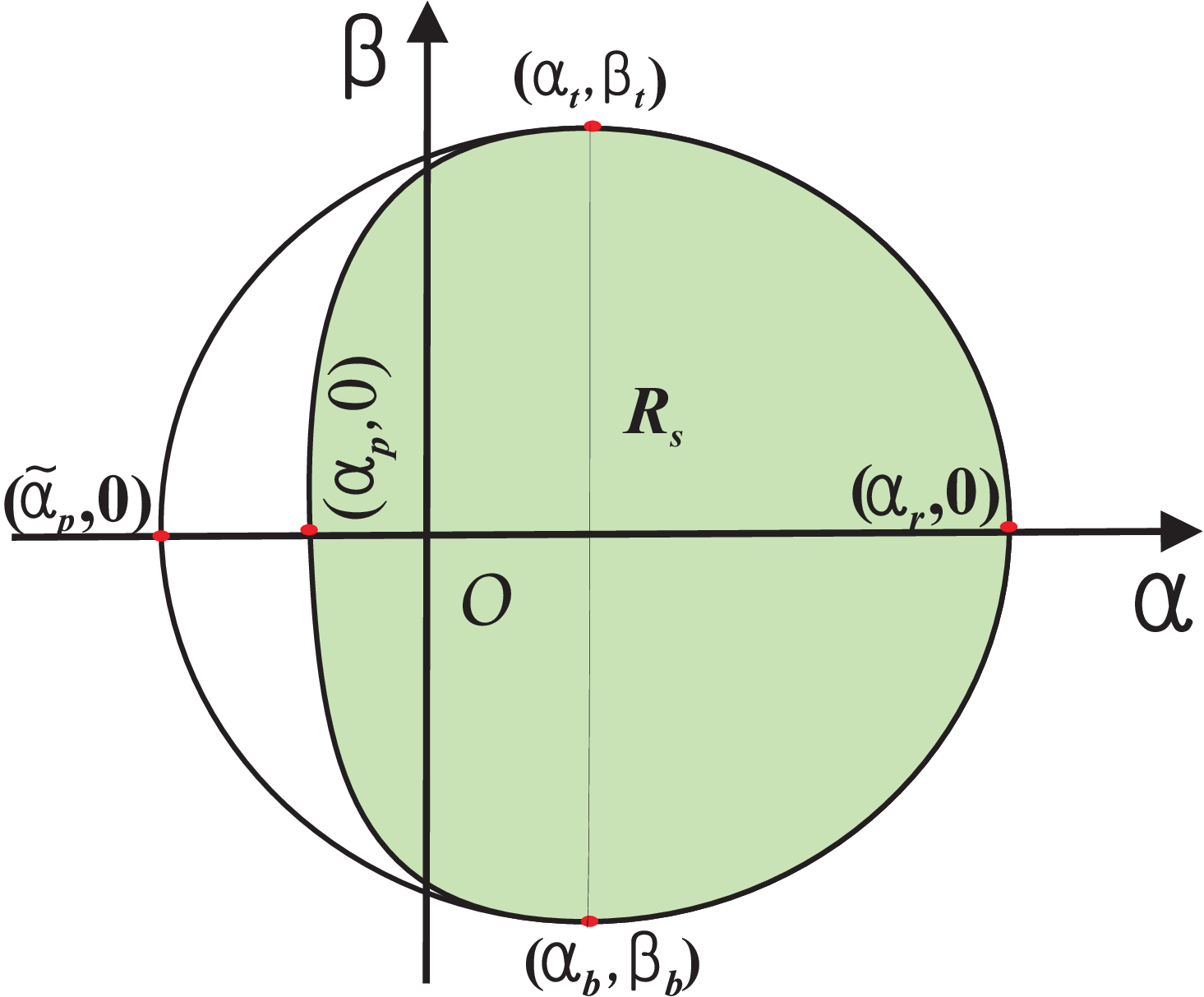} 
	 \end{tabular}
    \caption{\label{Fig3} Schematic representation of observable $R_s$ and $\delta_s$ ~\cite{Amir:2016cen}.}
    \end{center}
\end{figure*}
The motion of photons around the black hole can be parameterized by  the conserved quantities $\xi$ and $\eta$. For a non-rotating case ($a=0$) the shadow of the black hole is a perfect circle and can be obtained by the equation
\begin{equation}
\alpha^2+\beta^2=\frac{2 r_0 \left(-3 c^2 {r_0}^5+20 c M {r_0}^3-12 c q^2 {r_0}^2-12 M^2 r_0+8 M q^2+4 {r_0}^3\right)}{(r_0 (3 c r_0+2)-2 M)^2}=R_s^2.
\end{equation}
The contour of the above equation traces shadows of the non-rotating charged black hole with radius $R_s$, which we have plotted in Fig.~(\ref{Fig1}). The radius of non-rotating black hole shadow increases with the effect of quintessence while it can be observed from  Fig.~(\ref{Fig1}) that for the increasing values of charge, the size of the black hole shadow decreases.

For the rotating black hole $a\neq0$, one can find the equation  for celestial coordinates in terms of black hole parameters  as
\begin{eqnarray}\label{alphabeta}
{\alpha}^2+{\beta}^2&=&\frac{1}{(r_{0}(3cr_0+2)-2M)^2}\Big[ a^2 (r_0(2-3 c r_0)+2 M)^2\nonumber\\
&+&2 r_0 (-3 c^2 {r_0}^5+20 c M {r_0}^3-12 c q^2 {r_0}^2-12 M^2 {r_0}+8Mq^2+{4r_{0}}^{3}\Big]. \nonumber \\ 
\end{eqnarray}
The contour  of Eq.~(\ref{alphabeta}) for celestial coordinates $\alpha$ and $\beta$ traces the shadow of  rotating charged black hole in quintessence. For $q=0$, the above Eq. ~(\ref{alphabeta}) reduces for the Kerr black hole in quintessence\cite{Abdujabbarov:2015pqp}  
\begin{equation}
{\alpha}^2+{\beta}^2=\frac{a^2 ({r_0} (2-3 c {r_0})+2 M)^2-6 c^2 {r_0}^6+40 c M {r_0}^4-24 M^2 {r_0}^2+8 {r_0}^4}{({r_0} (3 c {r_0}+2)-2 M)^2}, \label{alphabeta2}
\end{equation}
and in the absence of quintessential energy, it reduces to the Kerr black hole \cite{Frolov:1998wf}. In Fig.~\ref{Fig2}, we plot the rotating charged black hole shadow in quintessence. The black hole shadow depends on its spin parameter $a$, normalization factor $c$, and charge $q$. In the absence of charge $q=0$, if we increase the values of normalization factor, we find that the size of the black hole shadow increases as shown by the first black color contour in all four plots in Fig.~\ref{Fig2} and also discussed in \cite{Abdujabbarov:2015pqp}. For fixed values of the spin parameter $a$ and normalization factor  $c$, the size of the black hole shadow decreases with charge $q$. The shape of the black hole shadow gets more and more distorted with increasing values of charge $q$. In comparison with the Kerr black hole shadow in quintessence \cite{Abdujabbarov:2015pqp}, the black hole shadow will appear smaller and more distorted with the effect of charge $q$.
\begin{figure*}
    \begin{tabular}{c c c c}
	\includegraphics[scale=0.6]{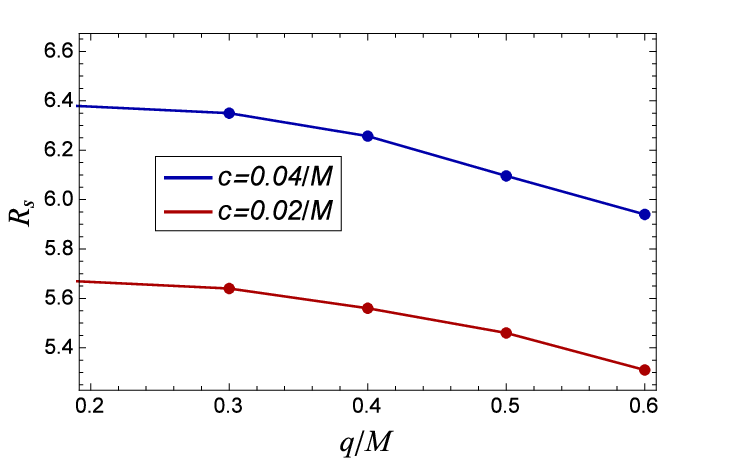} 
	\includegraphics[scale=0.6]{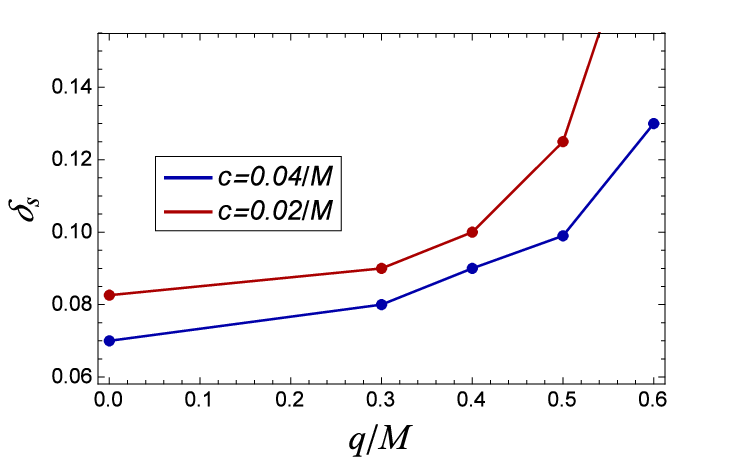} 
	 \end{tabular}
    \caption{\label{Fig4} Plot shows the variation of observable $R_s$ and $\delta_s$ for fix values of $a=0.8$ with charge $q$.}
\end{figure*}
For the analytical study of the shape and size of the black hole shadow contour,  we adopt two observables: the  shadow radius $R_s$ and the distortion parameter $\delta_s$ \cite{Hioki:2009na}. The study of these observable is totally based on the geometry of black hole shadow. The observable, shadow radius $R_s$ can be read as
\begin{equation}
R_s=\frac{({\alpha_t}-\alpha_r)^2+{\beta_t}^2}{2\mid{\alpha_t}-{\alpha_r}\mid},
\end{equation}
where $\alpha_r$ and $\alpha_t$ are the top and right positions on $\alpha$  axes and $\beta_t$ is the topmost position on the $\beta$-axes form where the reference circle passes (cf. Fig.{\ref{Fig3}). 
\begin{equation}
\delta=\frac{d}{R_s},
\end{equation}
where $d=\tilde{\alpha}_{p}-\alpha_{p}$ (cf. Fig.~\ref{Fig3}). 
 
In Fig.~\ref{Fig4} we have plotted  the observables shadow radius $R_s$ and distortion parameter $\delta_{s}$  for the increasing values of charge. It can be observed from Fig.~\ref{Fig4} that the observable shadow radius $R_s$ monotonically decreases while the observable distortion parameter  $\delta_s$  monotonically increases with the increasing values of charge $q$. In the absence of quintessential energy our results are equivalent to the Kerr--Newman black hole \cite{Takahashi:2005hy,Tsukamoto:2017fxq} and in the absence of charge the results approach to the Kerr black hole in quintessence  \cite{Abdujabbarov:2015pqp}. In comparison with the Bardeen black hole shadow in quintessence \cite{Rayimbaev:2022mrk}  we have found that the effect of magnetic charge $g$ on Bardeen black hole shadow in quintessence is small as compared to the effect of charge $q$ on rotating charged black hole shadow. In our case the size of black hole shadow decreases more rapidly as compared to the Bardeen black hole shadow in quintessence. Another interesting thing that we observed in our comparison is the magnetic charge $g$ does not much affect the distortion parameter, while in our case the black hole shadow gets effectively distorted with the charge parameter. A detailed comparison of shadow radius and distortion parameter with the charge parameter has shown in  Table~\ref{table1}.

 \begin{center}
 \begin{table}  
 
\begin{tabular}{|p{2cm}|p{2cm}|p{2cm}|}
 \hline
  \multicolumn{3}{|c|}{$c=0.02/M$} \\
 \hline
 charge parameter $(q/M)$& shadow radius $(R/M)$ & distortion parameter $(\delta_{s})$ \\
 \hline
0.0  & 5.72             &              0.08   \\
 0.3&   5.64   &             0.09   \\
0.4 &5.56 &                      0.10 \\
 0.5    &5.46       &              0.12   \\
  0.6   &5.31      &               0.20   \\
 
 \hline
\end{tabular}

\begin{tabular}{|p{2cm}|p{2cm}|p{2cm}|}
 \hline
  \multicolumn{3}{|c|}{$c=0.04/M$} \\
 \hline
 charge parameter $(q/M)$& shadow radius $(R/M)$ & distortion parameter $(\delta_{s})$ \\
 \hline
0.0  & 6.43             &               0.07   \\
 0.3&   6.35   &             0.08    \\
0.4 &6.25 &                       0.09 \\
 0.5    &6.09       &               0.10   \\
  0.6   &5.94      &               0.13   \\
 \hline

\end{tabular} 
\caption{Variation of shadow radius and distortion parameter with black hole charge.   } \label{table1}  
 \end{table}
 \end{center}
 
\section{Conclusion}\label{sect5}
In this work, we studied the optical properties of  rotating charged black holes  in the presence of quintessential dark energy.  
 The resulting properties of charged rotating black holes are affected by the presence of quintessence. We derived equations of motion of photon orbits and, on the basis of radial and angular equations, we calculated the impact parameters. For the visualization of the black hole's shadow, we derived the celestial coordinates in terms of impact parameters and plotted the contour of the black hole's shadow. In our study, we have found that  in the presence of quintessence, the size of  black hole shadow increases while the effect of charge monotonically decreases the radius of the black hole shadow. Another important result of our study is that the distortion parameter decreases with the normalization factor while the effect of charge monotonically increases the distortion parameter and the observed shadow gets more distorted with the increasing values of charge. 
\section{Acknowledgments}
 The author would like to thank Prof. S. G. Ghosh for his consistent support and special thanks to M. Amir for useful discussion.


\begin{thebibliography}{}
\bibitem{Akiyama:2019cqa}
  K.~Akiyama {\it et al.} [Event Horizon Telescope Collaboration],
 ``First M87 Event Horizon Telescope Results. I. The Shadow of the Supermassive Black Hole,''
  Astrophys.\ J.\  {\bf 875}, no. 1, L1 (2019).

\bibitem{Akiyama:2019brx}
  K.~Akiyama {\it et al.} [Event Horizon Telescope Collaboration],
 ``First M87 Event Horizon Telescope Results. II. Array and Instrumentation,''
  Astrophys.\ J.\  {\bf 875}, no. 1, L2 (2019).

\bibitem{Akiyama:2019sww}
  K.~Akiyama {\it et al.} [Event Horizon Telescope Collaboration],
 ``First M87 Event Horizon Telescope Results. III. Data Processing and Calibration,''
  Astrophys.\ J.\  {\bf 875}, no. 1, L3 (2019).

\bibitem{Akiyama:2019bqs}
  K.~Akiyama {\it et al.} [Event Horizon Telescope Collaboration],
 ``First M87 Event Horizon Telescope Results. IV. Imaging the Central Supermassive Black Hole,''
  Astrophys.\ J.\  {\bf 875}, no. 1, L4 (2019).

  \bibitem{Akiyama:2019fyp}
  K.~Akiyama {\it et al.} [Event Horizon Telescope Collaboration],
 ``First M87 Event Horizon Telescope Results. V. Physical Origin of the Asymmetric Ring,''
  Astrophys.\ J.\  {\bf 875}, no. 1, L5 (2019).


  \bibitem{Akiyama:2019eap}
  K.~Akiyama {\it et al.} [Event Horizon Telescope Collaboration],
  ``First M87 Event Horizon Telescope Results. VI. The Shadow and Mass of the Central Black Hole,''
  Astrophys.\ J.\  {\bf 875}, no. 1, L6 (2019).
  
  \bibitem{Synge:1966}
  J.L.~Synge,
  Mon.\ Not.\ R.\ Astron.\ Soc.\  {\bf 131}, 463 (1966).

\bibitem{Luminet:1979}
   J.P.~Luminet
  Astron.\ Astrophys. {\bf 75}, 228 (1979).


\bibitem{Bardeen:1973gb}
  J.~M.~Bardeen, in Black Holes, $Proceeding$ $of$ $the$ $Les$ $Houches$ $Summer$ $School$, $Session$ $215239$, edited by C. De Witt and B.S. De Witt and B.S. De Witt (Gorden and Breach, New York, 1973).
  
 
  \bibitem{Takahashi:2005hy}
  R.~Takahashi,
  Publ.\ Astron.\ Soc.\ Jap.\  {\bf 57}, 273 (2005).
  
  \bibitem{Wei:2013kza}
  S.~W.~Wei and Y.~X.~Liu,
  J.\ Cosmol.\ Astropart.\ Phys.\ 11 (2013) 063.
  
  
  \bibitem{Abdujabbarov:2012bn}
  A.~Abdujabbarov, F.~Atamurotov, Y.~Kucukakca, B.~Ahmedov, and U.~Camci,
  Astrophys.\ Space Sci.\  {\bf 344}, 429 (2013).
  
\bibitem{Amarilla:2011fx}
  L.~Amarilla and E.~F.~Eiroa,
  Phys.\ Rev.\ D {\bf 85}, 064019 (2012).

\bibitem{Amarilla:2013sj}
  L.~Amarilla and E.~F.~Eiroa,
  Phys.\ Rev.\ D {\bf 87}, 044057 (2013).
  
\bibitem{Atamurotov:2013sca}
  F.~Atamurotov, A.~Abdujabbarov, and B.~Ahmedov,
  Phys.\ Rev.\ D {\bf 88}, 064004 (2013).
  
\bibitem{Wang:2017hjl} 
  M.~Wang, S.~Chen and J.~Jing,
  JCAP {\bf 1710}, no. 10, 051 (2017).
  
\bibitem{sen:2009}
J. Schee and Z. Stuchlik,\ Int.\ Jour.\ Mod.\ Phys.\ D {\bf 18},
983 (2009).

\bibitem{Grenzebach:2014fha}
  A.~Grenzebach, V.~Perlick, and C.~L{\"a}mmerzahl,
  Phys.\ Rev.\ D {\bf 89}, 124004 (2014).

\bibitem{Amir:2016cen} 
  M.~Amir and S.~G.~Ghosh,
  Phys.\ Rev.\ D {\bf 94}, no. 2, 024054 (2016).
  
 
\bibitem{Abdujabbarov:2016hnw} 
  A.~Abdujabbarov, M.~Amir, B.~Ahmedov and S.~G.~Ghosh,
  Phys.\ Rev.\ D {\bf 93}, no. 10, 104004 (2016). 
 
\bibitem{Kumar:2019ohr}
R.~Kumar, B.~P.~Singh and S.~G.~Ghosh,
Annals Phys. \textbf{420}, 168252 (2020).

\bibitem{Kumar:2017tdw}
R.~Kumar, B.~P.~Singh, M.~S.~Ali and S.~G.~Ghosh,
Phys. Dark Univ. \textbf{34}, 100881 (2021).
 
\bibitem{Kumar:2020owy}
R.~Kumar and S.~G.~Ghosh,
JCAP \textbf{07}, 053 (2020).
  


  

  
  \bibitem{Atamurotov:2015nra} 
  F.~Atamurotov and B.~Ahmedov,
  Phys.\ Rev.\ D {\bf 92}, 084005 (2015).
  
\bibitem{Perlick:2015vta} 
  V.~Perlick, O.~Y.~Tsupko and G.~S.~Bisnovatyi-Kogan,
  Phys.\ Rev.\ D {\bf 92}, no. 10, 104031 (2015).
 
\bibitem{Goddi:2016jrs} 
  C.~Goddi {\it et al.},
  Int.\ J.\ Mod.\ Phys.\ D {\bf 26}, no. 02, 1730001 (2016).
    
  
    
\bibitem{Cunha:2015yba} 
  P.V.~P.~Cunha, C.A.~R.~Herdeiro, E.~Radu, and H.F.~Runarsson,
  Phys.\ Rev.\ Lett.\  {\bf 115}, 211102 (2015).
  
\bibitem{Cunha:2016bpi} 
  P.~V.~P.~Cunha, C.~A.~R.~Herdeiro, E.~Radu and H.~F.~Runarsson,
  Int.\ J.\ Mod.\ Phys.\ D {\bf 25}, no. 09, 1641021 (2016).
 
\bibitem{Hioki:2009na}
K.~Hioki and K.~i.~Maeda,
Phys. Rev. D \textbf{80}, 024042 (2009).

\bibitem{Abdujabbarov:2015xqa} 
  A.A.~Abdujabbarov, L.~Rezzolla, and B.~J.~Ahmedov,
  Mon.\ Not.\ R.\ Astron.\ Soc.\  {\bf 454}, 2423 (2015).  

\bibitem{Kumar:2018ple}
R.~Kumar and S.~G.~Ghosh,
Astrophys. J. \textbf{892}, 78 (2020).


\bibitem{Chang:2020miq}
Z.~Chang and Q.~H.~Zhu,
Phys. Rev. D \textbf{101}, no.8, 084029 (2020).

\bibitem{Pesce:2021adg}
D.~W.~Pesce, D.~C.~M.~Palumbo, R.~Narayan, L.~Blackburn, S.~S.~Doeleman, M.~D.~Johnson, C.~P.~Ma, N.~M.~Nagar, P.~Natarajan and A.~Ricarte,
Astrophys. J. \textbf{923}, no.2, 260 (2021).

\bibitem{Papnoi:2014aaa}
  U.~Papnoi, F.~Atamurotov, S.~G.~Ghosh, and B.~Ahmedov,
  Phys.\ Rev.\ D {\bf 90}, 024073 (2014).
  
\bibitem{Abdujabbarov:2015rqa} 
  A.~Abdujabbarov, F.~Atamurotov, N.~Dadhich, B.~Ahmedov and Z.~Stuchlík,
  Eur.\ Phys.\ J.\ C {\bf 75}, no. 8, 399 (2015).
  
\bibitem{Amir:2017slq}
M.~Amir, B.~P.~Singh and S.~G.~Ghosh,
Eur. Phys. J. C \textbf{78}, no.5, 399 (2018).
  
\bibitem{Singh:2017vfr}
B.~P.~Singh and S.~G.~Ghosh,
Annals Phys. \textbf{395}, 127-137 (2018).


\bibitem{Belhaj:2020rdb}
A.~Belhaj, M.~Benali, A.~El Balali, H.~El Moumni and S.~E.~Ennadifi,
Class. Quant. Grav. \textbf{37}, no.21, 215004 (2020).

\bibitem{SupernovaSearchTeam:2004lze}
A.~G.~Riess \textit{et al.} [Supernova Search Team],
Astrophys. J. \textbf{607}, 665-687 (2004).

\bibitem{Peebles:2002gy}
P.~J.~E.~Peebles and B.~Ratra,
Rev. Mod. Phys. \textbf{75}, 559-606 (2003).



\bibitem{Copeland:2006wr}
E.~J.~Copeland, M.~Sami and S.~Tsujikawa,
Int. J. Mod. Phys. D \textbf{15}, 1753-1936 (2006).


\bibitem{WMAP:2010qai}
E.~Komatsu \textit{et al.} [WMAP],
Astrophys. J. Suppl. \textbf{192}, 18 (2011).

\bibitem{Kiselev:2002dx}
V.~V.~Kiselev,
Class. Quant. Grav. \textbf{20}, 1187-1198 (2003).

\bibitem{Abdujabbarov:2015pqp}
A.~Abdujabbarov, B.~Toshmatov, Z.~Stuchl\'\i{}k and B.~Ahmedov,
Int. J. Mod. Phys. D \textbf{26}, no.06, 1750051 (2016).

\bibitem{Belhaj2022}
 A.~Belhaj and Y.~Sekhmani
Z.~Xu and J.~Wang,
General Relativity and Gravitation  \textbf{54}, no. 17 (2022).

\bibitem{Zeng:2020vsj}
X.~X.~Zeng and H.~Q.~Zhang,
Eur. Phys. J. C \textbf{80}, no.11, 1058 (2020).

\bibitem{Sun:2022wya}
C.~Sun, Y.~Liu, W.~L.~Qian and R.~Yue,
[arXiv:2201.01890 [gr-qc]].

\bibitem{Pedraza:2020uuy}
O.~Pedraza, L.~A.~L\'opez, R.~Arceo and I.~Cabrera-Munguia,
Gen. Rel. Grav. \textbf{53}, no.3, 24 (2021).




\bibitem{Xu:2016jod}
Z.~Xu and J.~Wang,
Phys. Rev. D \textbf{95}, no.6, 064015 (2017).


\bibitem{Newman-Janis:1965}
Newman, E.T. and Janis, A. I. 1965, Journal of Mathematical Physics, 6, 915.


\bibitem{Azreg-Ainou:2014pra}
M.~Azreg-A\"\i{}nou,
Phys. Rev. D \textbf{90}, no.6, 064041 (2014).

\bibitem{Ghosh:2015ovj}
S.~G.~Ghosh,
Eur. Phys. J. C \textbf{76}, no.4, 222 (2016).

\bibitem{Toshmatov:2015npp}
B.~Toshmatov, Z.~Stuchl\'\i{}k and B.~Ahmedov,
Eur. Phys. J. Plus \textbf{132}, no.2, 98 (2017).

\bibitem{Carter:1968rr} 
  B.~Carter, 
  Phys.\ Rev.\ {\bf 174}, 1559 (1968).
  

\bibitem{Frolov:1998wf}
V.~P.~Frolov and I.~D.~Novikov,
doi:10.1007/978-94-011-5139-9.



\bibitem{Tsukamoto:2017fxq}
N.~Tsukamoto,
Phys. Rev. D \textbf{97}, no.6, 064021 (2018).

\bibitem{Rayimbaev:2022mrk}
J.~Rayimbaev, B.~Majeed, M.~Jamil, K.~Jusufi and A.~Wang,
Phys. Dark Univ. \textbf{35}, 100930 (2022).

 \end{thebibliography}
\end{document}